\documentclass[12pt]{iopart}
\usepackage{graphicx}
\usepackage{xcolor} 

\begin{document}

\title[Graphdiynes interacting with metal surfaces]{Graphdiynes interacting with metal surfaces: first-principles electronic and vibrational properties}

\author{Simona Achilli$^{a,b,c}$, Alberto Milani$^{d}$, Guido Fratesi$^{a,b,c}$, Francesco Tumino$^{d}$, Nicola Manini$^{a,b}$, Giovanni Onida$^{a,b,c}$, Carlo S. Casari$^{d,*}$}

\address{
$^{a}$Dipartimento di Fisica ``Aldo Pontremoli'', 
$^{b}$European Theoretical Spectroscopy Facility (ETSF).
$^{c}$INFN, Sezione di Milano, I-20133 Milano, Italy.
$^{d}$Micro and Nanostructured Material Lab - NanoLab, Dipartimento di Energia, Politecnico di Milano, Via Ponzio 34/3 I-20133 Milano, Italy
}

\ead{$^{*}$carlo.casari@polimi.it}
\vspace{10pt}
\begin{indented}
\item[]
\end{indented}

\begin{abstract}
Graphdiynes (GDYs) represent a class of 2D carbon materials based on sp-sp$^2$ hybridization with appealing properties and potential applications. 
Recent advances have demonstrated the experimental 
self-assembly
of GDYs on metal substrates.
Here we focus on $\alpha$- and $\beta$-GDYs on Au(111) and Pt(111), 
and 
investigate how their electronic and vibrational properties are affected by the interaction with a metal substrate.
We adopt hydrogenated GDY, previously characterized experimentally,
as a benchmark for density functional theory simulations, that we apply to show that Au and Pt substrates impose a different degree of distortion on both $\alpha$- and $\beta$-GDY.
By comparing the adsorbed and the freestanding structures, we evaluate the effect of the surface interaction
on the bandstructure and the simulated Raman spectra.
Different charge transfers result in different energy shift of the Dirac cone in semi-metallic $\alpha$-GDY and changes from semiconducting to metallic behavior for $\beta$-GDY.
These changes in electronic properties are accompanied by characteristic frequency shifts and modifications of Raman active modes.
Our results contribute in the understanding of the metal-interaction effects on GDYs and can open
a route
to the design of novel 2D materials with tailored properties.           
\end{abstract}

%
\vspace{2pc}
\noindent{\it Keywords}: Graphdiynes, DFT, Raman, sp-carbon, 2D carbon structures
%

%
%
%

\section{Introduction}

Graphdiynes (GDYs) are receiving increasing interest as a new class of 2D carbon materials \cite{Hirsch_NatMat_2010,Sakamoto-AdvMat2019-rev, Liu21, Zou20,Yu21}. Different GDY structures can be conceived by playing with sp$^2$ and sp hybridized carbon atoms \cite{baughman1987,Huang_2018,Casari_Nanoscale_2016,Casari_MRSComm_2018}.
For instance, starting from graphene and conceptually
substituting sp$^2$ carbon atoms with diacetylenic linear units as linkages, a total of 26 possible novel 2D carbon crystals (with up to 8 atoms per unit cell) can be formed with variable sp/sp$^2$ ratio, porosity and density \cite{Serafini-Proserpio-JPCC2021}. 
Such 2D materials exhibit
electronic properties ranging from metallic to semi-metallic with Dirac cones and to semiconducting with different energy gap, making them appealing for technological applications\cite{Gao_2019, ChenZhi-AnnDerPhys2017,Jia_2017,Li_NanoEn_2018,Zuo_2018}.
The experimental realization of GDY is quite recent and it is mainly related to
the so-called $\gamma$-GDY showing interesting performances for photocatalysis, catalysis, photovoltaics and electronic device realization \cite{Li_2010_GDY,Gao_2019,Li_ChemSocRev_2014}.
In this framework, Raman spectroscopy is one of the key techniques for the characterization of nanostructured carbon materials, including GDY-based systems \cite{Ivanovskii_ProgSolidStChem_2013,serafini2020raman,Zhang_JPCC2016-graphyne}.
For its particular sensitivity to the local molecular bond, hybridization state, and to the structural order, Raman spectroscopy is
currently
a fundamental tool to recognize the fingerprints  of specific sp-carbon based materials with respect to other carbon nanostructures \cite{Milani_BeilsteinJ_2015}. 

The synthesis approach for GDYs is mainly based on the use of reactive molecular building blocks which assemble in a 2D structure, usually catalyzed by the presence of a metal \cite{zhang2012homo}. Such approach, first developed in the liquid phase, can also be used in vacuum exploiting on-surface synthesis techniques.
When the on-surface synthesis is performed in ultra high vacuum on atomically flat metal surfaces (e.g., Au(111), Cu(111), Pt(111)),  a direct atomic scale characterization by means of surface science techniques such as scanning tunneling microscopy (STM) becomes possible \cite{Sun_JACS_2016, Sun_AngewChem_2017, sun2016, Fan20152484,klappenberger2015surface}. Different works used STM to study on-surface synthesized sp-sp$^2$ carbon structures with impressive imaging capabilities, showing the formation process down to the atomic scale and visualizing the carbon-carbon bonds \cite{Shu_NatComm_2018}.    

The specific metal substrate suitable for GDY synthesis (e.g., copper, gold, platinum, silver) plays a complex role in modifying the carbon overlayer properties through different phenomena, including strain to adapt to the metal surface lattice parameters, charge transfer, and even partial hybridization of carbon atoms with the metal substrate ones.
For instance, the $\gamma$-GDY metal contact has been recently theoretically investigated in view of a field effect transistor realization \cite{Pan}.

The case of hydrogenated GDY systems (h-GDYs) has been studied by means of a combined Raman spectroscopy and STM investigation \cite{rabia2020ANM}. 
The theoretically computed electronic and vibrational properties for an Au(111)-synthesized h-GDY system turned out to be consistent
with its experimentally measured properties.
In addition,
the theoretical analysis naturally allows one to compare the surface-deposited case with the freestanding GDY, outlining the importance of interaction effects with the metal surface. Even though recent works have shown the possibility to exfoliate GDY from multilayer samples and to transfer flakes to insulating substrates, \cite{Yan} the presence of a metal substrate is usually required in the synthesis step so that the majority of GDY systems experimentally investigated so far is interacting with or supported on a metal.       

Hence, we focus here on the so-far poorly explored
possibilities provided by metal substrates to tune the GDY properties as an additional versatility for material engineering. In this work, we analyze two different GDY structures on two typical metals, namely $\alpha$-GDY and $\beta$-GDY on Au(111) and Pt(111), as representative substrates capable of a weak and 
relatively strong interaction with GDYs, respectively. The freestanding $\alpha$-GDY and $\beta$-GDY display dissimilar electronic properties: $\alpha$-GDY is a semimetal, with a Dirac cone at the Fermi level and graphene-like band structure, while $\beta$-GDY is a semiconductor with a direct gap of 0.26~eV, 
obtained in the present work by means of density-functional theory (DFT) with the Perdew-Burke-Ernzerhof (PBE) exchange-correlation functional.
$\alpha$- and $\beta$-GDYs and their interaction with the supporting metal surfaces are poorly
investigated in the literature which focuses mainly on the
$\gamma$-GDY \cite{Li_ChemSocRev_2014,serafini2020raman} and on similar available systems, also studied in recent experimental works \cite{rabia2019scanning,rabia2020ANM, Borrelli-Angwchemie2021, Galeotti-NatMat2020}.

Our theoretical analysis focuses on the effects of the interaction between GDY and the underlying metals, in term of structural deformation, charge transfer, and changes in the electronic band structure. The study includes a detailed analysis of the consequent effects on the GDY's vibrational properties. 
In particular,
we investigate how the overlayer-substrate interaction modifies the electronic properties at the Fermi level, with particular attention to the changes induced on the Dirac cone in the 
semi-metallic
$\alpha$-GDY case, and to the possibility to turn the semiconductor $\beta$-GDY into a metallic system.
Due to the changes in the electronic structure, and in view of the  strong electron-phonon coupling present in these materials \cite{Casari_MRSComm_2018, Ivanovskii_ProgSolidStChem_2013}, substantial effects are to be expected in the Raman spectra.
Indeed, these spectra turn out to be strongly sensitive to the interaction with the substrate, and can be used to monitor and to predict the GDY structural changes due to the interaction with the metal surface.

On the other hand, vibrational properties, and in particular  the Raman response, of these sp-sp$^2$ hybrid materials have been only rarely investigated by DFT-based theoretical works. It is indeed often computationally prohibitive to deal with extended systems characterized by a large periodicity imposed by the matching between the carbon network and the substrate, in the presence of metallicity. For this reason, simplified models
are commonly adopted in such simulations
to account for the interaction with the underlying substrate.

In Refs. \cite{rabia2019scanning,rabia2020ANM}
Raman spectra were computed with the simplifying assumption of a model formed by a fragment of the full system, interacting with small Au clusters. That model provided a reliable interpretation of the experimental Raman spectra of both 1D and 2D systems, including the spectral changes observed upon interaction with the gold substrate, shedding light on the molecular phenomena involved. However, the choice of the fragment and the size and position of clusters must be carefully gauged, to avoid results dependent on the selected model.
Here we propose and test a more general and less arbitrary scheme, accounting for the full lateral extension of the GDY, and keeping into account the substrate effects through the structural deformation induced on the GDY by the interaction with the metal substrate,
although the last one is
not explicitly included in the calculation of the Raman spectra.
After successfully validating such an approach for the h-GDY test case, we turn our attention on the novel GDY systems on different metals, focusing on the study of the role of the GDY-substrate interaction in determining the vibrational properties and Raman spectra.

\section{Methods}

We performed the on-surface synthesis of h-GDY in ultra high vacuum (base pressure of less than $5\times 10^{-11}$~mbar. The Au(111) surface was prepared by repeated cycles of Ar$^+$ ion sputtering, followed by annealing at 720~K. By means of an organic molecular
evaporator (OME) at 304~K, we evaporated
the molecular precursor (1,3,5-tris(bromoethynyl)benzene (tBEP)) on the cleaned Au(111) surface kept
at room temperature (RT).

When deposited on Au, terminal Br atoms in tBEP are released to form intermolecular connections through Au adatoms.
At room temperature an organo-metallic system is formed, displaying the 2D network driven by the trigonal geometry of the precursor molecule. 
A subsequent thermal annealing at 480~K for about half an hour promotes the removal of Au adatom to form diacetylenic units by homocoupling reactions.
Thermal treatment at higher temperatures leads to increased disorder in the network.       
We performed room-temperature STM measurements by means of
an Omicron variable-temperature scanning tunneling microscope. STM images were taken in constant-current mode, with a chemically etched tungsten tip.
We conducted ex-situ Raman measurements using a Renishaw InVia spectrometer coupled with an Argon laser (514.5~nm) with laser power of 5~mW.

We carry out DFT calculations with the generalized gradient approximation (GGA) in the PBE form
\cite{PBE}, and including
van der Waals interactions between the organic overlayer and the substrate via a DFT-D2
Grimme potential \cite{Grimme}.
We use the approach implemented in the SIESTA code \cite{Sole02} that relies on norm-conserving pseudopotentials and an atomic-orbitals
basis set.  This method allows us to treat
fairly large unit cells, such as those formed by matching the considered GDY structures with the substrates, namely a $4 \times 4$ and a $5 \times 5$ supercell (see section \ref{sec3.2}). Moreover, it provides a first-principles characterization of the structural and electronic properties of the systems induced by the interaction with the substrate and allows us to generate simulated STM images using the Tersoff-Hamann approach \cite{Tersoff}.

We adopt a double-zeta basis set with polarization orbitals, a mesh cutoff for the real space grid of
400~Ry, and a $5 \times 5$ and $4 \times 4$ sampling of the Brillouin zone for the $4 \times 4$ and $5 \times 5$ surface reconstructions, respectively.
A five times denser $\mathbf k$-space
grid is used for the calculation of the density
of states (DOS). We relax the organic layers and the first substrate layer until the forces reach the tolerance value of 0.04~eV\,\AA$^{-1}$.
The details of the calculation for h-GDY on Au(111) are reported in Ref. \cite{rabia2020ANM}.

We compute the Raman spectra by DFT in a periodic boundary conditions (PBC) approach  keeping a Gaussian basis set (GBS) (as implemented in the CRYSTAL package). 
Since DFT-PBC calculations of Raman intensities on a Gaussian basis sets cannot be adopted to describe metal surfaces, due to unavoidable convergence problems in the simulations, the computation of the Raman spectra have been carried out on the optimized crystal geometries of GDYs only, both in the free-standing case and for GDYs in the distorted configuration, as resulting from the interaction with Au and Pt surface.
These configurations are the optimized geometries as determined by the SIESTA calculations discussed above (keeping the same GGA functional also for the Raman calculations).
By this procedure, it is thus possible to simulate the structural effects on the vibrational spectra of a 2D extended system as a consequence of the interaction with metal surfaces, avoiding more arbitrary and size-restricted molecular models formed by finite-size fragments interacting with small Au clusters.

As a benchmark for the adopted method, Fig.~S4 of
the Supporting Information reports the comparison of the Raman spectra obtained by DFT-PBC-GBS calculations of the free-standing systems here considered (namely, h-GDY, $\alpha$-GDY, and $\beta$-GDY) after full geometry optimization.
The comparison involves the PBE functional adopted for most calculations here, and the PBE0 hybrid functional\cite{Adamo-BaroneJPC1999-PBE0}, which is considered the standard reference for Raman spectra of molecular systems \cite{TommasiniJPCA2007,Milani_BeilsteinJ_2015}.

These two spectra are in agreement, presenting a very similar pattern in wavenumbers and intensities, confirming the reliability of the computational approach here adopted.

\section{Results and Discussion}

\subsection{Hydrogenated graphdiyne(h-GDY) on Au(111) as an experimentally available system: theory and experiment}

\begin{figure}
  \centering
  \includegraphics[width=0.8\linewidth]{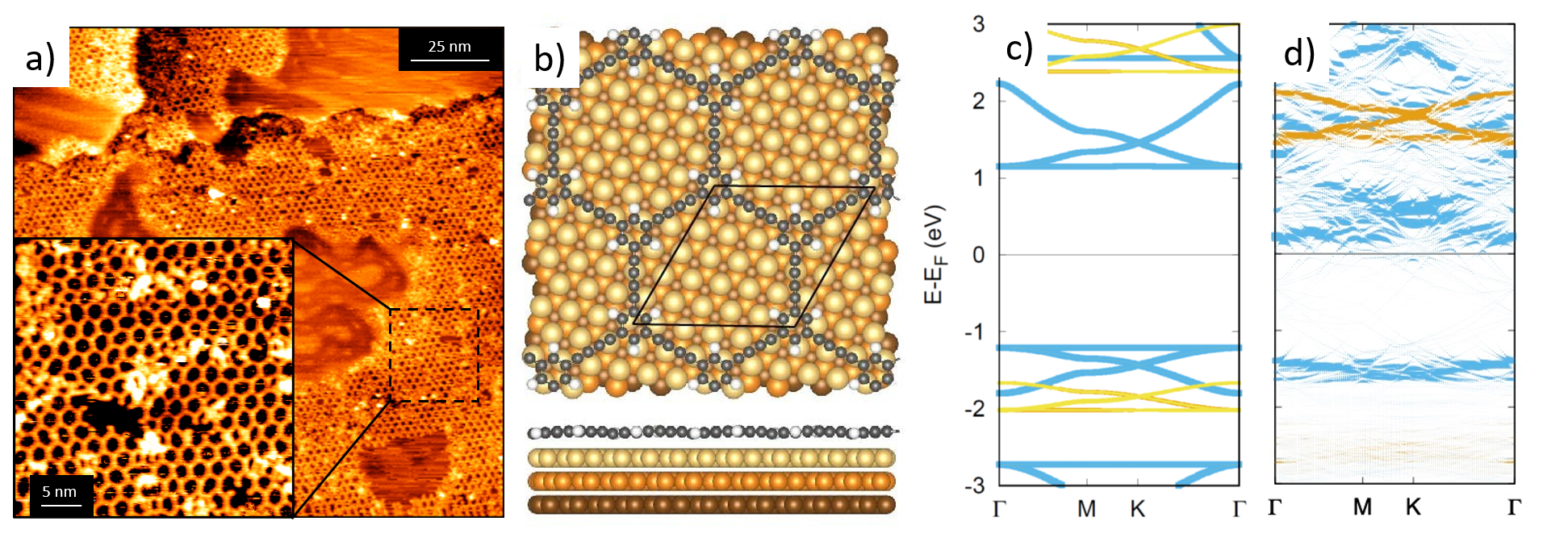}
  \caption{(a) STM images of h-GDY on Au(111) produced by on-surface synthesis.
  Inset: a magnification resolving the structure. Both images are collected at $-0.6$~V and $0.3$~nA.
(b) Top and side view of our model for h-GDY on Au(111); the black line highlights the periodically repeated cell; this model was first adopted in \cite{rabia2020ANM}.
  (c) Band structure of the freestanding and (d) supported h-GDY. In panel (d) the states in blue and yellow/orange have p$_z$ (orthogonal to the surface plane) and p$_x$/p$_y$ character, respectively.}
  \label{fig:struct-hGDY}
\end{figure}

In Ref. \cite{rabia2020ANM} a multi-level characterization of the properties of h-GDY on Au(111) based on structure investigation by STM and Raman spectroscopy was reported. 
We measured STM images on this system, reported in Figure~\ref{fig:struct-hGDY}a, showing the formation of an extended hexagonal structure on top of Au(111) surface.
Figure \ref{fig:struct-hGDY}b reports
the structural model, which
is characterized by a 9$^\circ$ rotation of the h-GDY network relative to the $[11\bar{2}]$ surface direction.
This same model could
reproduce the experimental STM images \cite{rabia2020ANM}.
The structural relaxation induced by the interaction with gold results in a mild bending of h-GDY
relative to the freestanding case.

Concerning the electronic properties, the band structure
(Fig.\ref{fig:struct-hGDY}c,d) 
shows the downward energy shift of the 
empty
p-states of the h-GDY, with a dramatic reduction of the electronic gap
compared
to the freestanding system (Fig.\ref{fig:struct-hGDY}c).
Even though
the energy gap between p$_z$ bands
remains
visible, the charge transfer from the Au(111) surface to the h-GDY turns the carbon network into a 
weakly
metallic system, with originally empty p$_z$ states downshifted in energy and
touching 
the Fermi level.

\begin{figure}
  \centering
  \includegraphics[width=0.8\linewidth]{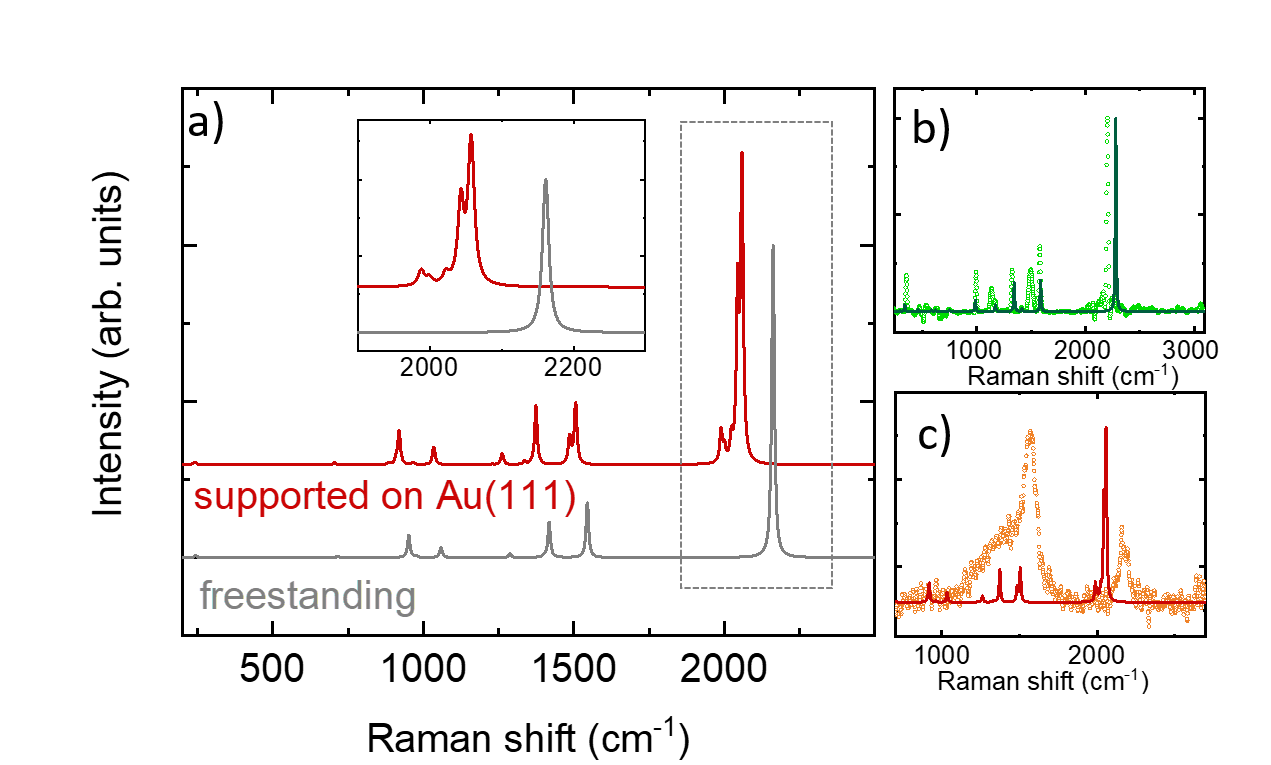}
  \caption{(a) DFT simulated Raman spectra of freestanding h-GDY and h-GDY on Au(111).
  Inset: a close-up of the spectral region of the ECC mode.
    (b) Comparison of DFT and experimental Raman spectra of the precursor molecule (tBEP).
  (c) Comparison of DFT and experimental Raman spectra of h-GDY system on Au(111).
  }
  \label{fig:Raman-hGDY}
\end{figure}

Figure \ref{fig:Raman-hGDY}a
reports the Raman spectra for the h-GDY  system, carried out by using the approach presented above, for both the freestanding system and for h-GDY supported on Au(111).
DFT simulations tend to systematically overestimate 
the conjugation giving different peak frequencies compared to experiment, by a roughly constant factor.
It is a standard and widely adopted method to rescale the simulated frequencies using a scaling factor evaluated by comparison with experimental vibrational frequencies.
However, Figure~\ref{fig:Raman-hGDY}b
shows a very good agreement
in the comparison of the experimental and simulated spectra, in particular the almost coinciding frequency of the stretching mode of the phenyl ring at 1581~cm$^{-1}$, indicating no need of any rescaling.
Hence all the simulated Raman spectra reported in this work are unscaled.

When h-GDY interacts with the Au surface, peculiar effects are found in the Raman spectra. In particular, the peak above 2000~cm$^{-1}$, assigned to the collective CC stretching of the sp-domains (the so-called Effective Conjugation Coordinate –  ECC mode of polyynes \cite{Milani_BeilsteinJ_2015,Casari_Nanoscale_2016}), 
exhibits an evident downshift upon interaction with Au
compared to the freestanding case. 
In addition to a frequency shift, new features appear in the region of the ECC band when the system is supported on Au(111), relative to the single peak
observed for the freestanding system.
This single peak is 
in fact the convolution of three contributions associated to different Raman-active combination of the ECC modes on the different diacetylenic units in the cell, almost coincident in wavenumber.
Conversely the h-GDY interacting on Au 
exhibits distinctive features 
between 1980 and 2060 cm$^{-1}$ of which three are again the most intense ones (2022, 2043, and 2057 cm$^{-1}$, see inset of Fig.\ref{fig:Raman-hGDY}a).
Inspection of the normal modes reveals that these most intense bands are still related to collective ECC vibrations on the diacetylenic units, while the weaker bands at lower wavenumbers (1988 and 2000 cm$^{-1}$) are CC stretching vibrations more localized on specific bonds 
in the sp-carbon domains. 
This is in agreement with previous works
in which also the pattern or normal modes was found to be affected by the specific interaction with gold.
In particular, the collective ECC mode on the diacetylenic 
units changes into more localized CC stretching vibrations. 
At frequencies below 1600~cm$^{-1}$, vibrations of the aromatic units are found, but they are not so significantly influenced by the interaction with the gold surface.

In previous works \cite{rabia2019scanning,rabia2020ANM}
the observed frequency downshift of the ECC mode has been interpreted  as the result of the strong interaction with Au which is responsible for weakening the CC bonds, thus lowering  their stretching force constant and causing lower vibrational wavenumbers.
In the case of finite-length sp-carbon wires (i.e., polyynes), a downshift of the ECC mode is usually correlated to a more 
equalized structure (i.e., a smaller bond-length alternation -- BLA -- between the quasi-single and quasi-triple CC bonds) and also a smaller band gap, or as a consequence of tensile strain \cite{Castelli2012}.
By computing the BLA of the di-acetylenic units in the h-GDY network (as the difference between the average bond lengths of the single CC bonds and the bond length of the triple CC bond at the center of the $sp$-carbon chain), values of 0.123~\AA\ and 0.119~\AA\ are found for the freestanding and  Au-supported systems, respectively.
The difference between these two values (0.004~\AA) is however not
sufficient to justify the downshift observed in the Raman spectra. 
Analyzing the value of the length of the different CC bond in the sp-carbon units, in the Au-supported system 
longer values are found, with all bonds longer by  0.01--0.02~\AA\ than in the free-standing case.
The interaction with gold hence reduces the bond strength of the CC bonds in the sp-carbon units, thus implying softer stretching force constants and justifying the downshift in wavenumber of the related stretching vibrational modes. These results are consistent with those discussed in
\cite{rabia2019scanning, rabia2020ANM}, using DFT on molecular fragments interacting with small Au clusters. 
The spectral changes found in previous investigations also indicated a change of normal modes upon interaction with gold, namely the formation of more localized CC stretching vibrations induced by the presence of gold. This led to the appearance of peaks well below 2000~cm$^{-1}$
(1939 and 1969~cm$^{-1}$),
which are hardly observed in the experimental spectrum \cite{rabia2020ANM}.
The main features of the experimental Raman spectrum are well reproduced by the current computational approach (see Fig.\ref{fig:Raman-hGDY}c).
The splitting of the ECC mode into many different peaks all in a small spectral range can account for the
broad band observed in the experiment.
However, a mismatch is present when considering the experimental and simulated spectrum of h-GDY on Au(111). The slightly different peak frequency of ECC mode in simulated and experimental spectra can be related to the well-known difficulties of DFT in precisely estimating the degree of conjugation. 
In addition, the experimental spectrum shows a broad and intense band in the sp$^2$-carbon region below 1600~cm$^{-1}$, which is reasonable considering the presence of defects and disorder, not included in the simulation and giving rise to the so-called G and D features typical of sp$^2$ amorphous carbon.

\subsection{$\alpha$- and $\beta$-graphdiynes on metals}

\subsubsection{Structural models and electronic properties.}
\label{sec3.2}

\begin{figure}
  \centering
  \includegraphics[width=1.0\linewidth]{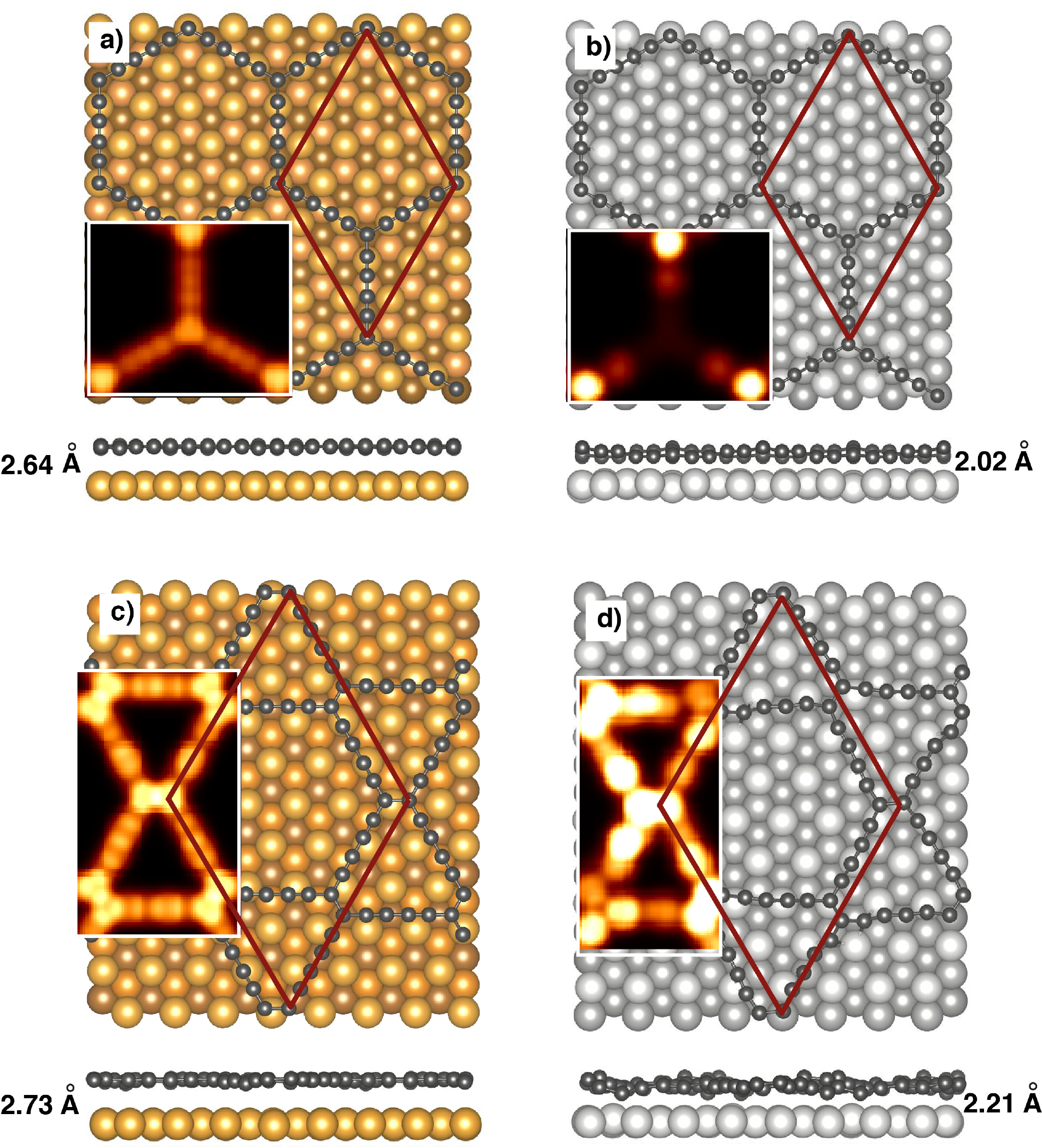}
  \caption{Top view and 
    side view (two outermost layers only) of (a) $\alpha$-GDY on Au(111), (b)  $\alpha$-GDY on Pt(111), (c) $\beta$-GDY on Au(111), and (d) $\beta$-GDY on Pt(111).
  The average distance between the GDY and the surface layer is indicated, and the unit cell is marked in red. Atoms in the first-, second- and third- substrate layer are marked with different shades.
  The simulated STM images obtained for filled states (integration range: from 0.5~eV below the Fermi energy $\epsilon_{\rm F}$ up to $\epsilon_{\rm F}$;
  tip distance: 2~{\AA} from the GDY average plane) are reported as aligned insets. 
  The color scale of STM images is the same for all the panels, ranging from 0.0 to 0.0035 e/Bohr$^3$.
 }
  \label{fig:struct}
\end{figure}

The PBE-relaxed freestanding structures of $\alpha$-GDY and $\beta$-GDY exhibit
periodicities of 11.493~{\AA} and 14.704~{\AA}, respectively (slightly larger than obtained in PBE0, namely 11.398~\AA\ and 14.600~\AA, respectively). 
They can be matched with the same supercell for both substrates, given their similar lattice constants: 4.21~{\AA} for Au and 4.18~{\AA} for Pt in our PBE model.
In particular, we adopt a structural model in which $\alpha$-GDY is matched to a $4 \times 4$ supercell of the substrate and $\beta$-GDY to a $5 \times 5$ one.
The resulting strain of the GDY structures is 3.6\% (2.9\%) and 1.2\% (0.5\%) for  $\alpha$-GDY ($\beta$-GDY) on Au(111) and Pt(111), respectively.
The sp-carbon linking chains
of the $\alpha$-GDY are aligned along the $[11\bar{2}]$ directions of the two surfaces (see Figure \ref{fig:struct}a,b), while in the $\beta$-GDY they follow the surface $[1\bar{1}0]$ directions and appear rotated by 30$^\circ$
relative
to the former (see Fig.~\ref{fig:struct}c,d).

Both $\alpha$-GDY and $\beta$-GDY show a
fairly
small structural relaxation on Au(111), contrasted with the remarkable distortion on Pt(111), due to a stronger interaction with the Pt substrate.
In particular, $\alpha$-GDY remains almost flat on Au(111) (the spread in the $z$ coordinate, orthogonal to the surface, is 0.18~\AA), with an average distance of 2.64~{\AA} from the outermost surface layer, while on Pt(111) the GDY is more 
corrugated (0.63~\AA), and lies at a $\simeq$ 23\%
smaller average distance from the surface (2.02~\AA).\footnote{
The average distance is obtained 
as the difference between
the mean $z$ coordinate of the GDY and of the 
uppermost
surface layer.
The corrugation is determined as the difference between the maximum and the minimum value of the $z$ coordinates in the overlayer.}

This difference between the two substrates is confirmed also in the $\beta$-GDY case. 
The carbon network, although slightly more corrugated than 
the $\alpha$-GDY system, is still  almost flat on Au(111), with a corrugation of 0.31~\AA.
When deposited on Pt(111) the $\beta$-GDY structure becomes instead largely distorted, both in-plane and out-of-plane: we observe a deformation of the $sp$-carbon units and an overall vertical corrugation of 0.98~\AA.
By comparing $\alpha$-GDY with $\beta$-GDY one can note that the latter sits slightly farther from the surface (average distance 2.73~{\AA} on Au(111) and 2.21~{\AA} on Pt(111)), suggesting a weaker interaction with the substrate. 

The larger distortion of the $\beta$-GDY
compared
to the $\alpha$-GDY can be understood in terms of the different orientation of the $sp$-chains
relative
to the underlying substrates.
Indeed in 
$\beta$-GDY
the interaction with atoms of the top surface layer is favored by the alignment along the $[1\bar{1}0]$ crystalline lines.

The insets of Fig.~\ref{fig:struct}
report the simulated STM  images of the occupied states (bias $V=-0.5$~V).
These four images, 
taken at the same height above the average $z$ of the organic layer and
reported in the same color scale, 
exhibit
a different contrast on the GDY layer, depending on the substrate on which they are adsorbed.
In $\alpha$-GDY/Au the whole hexagonal structure of the GDY is visible, including contributions also from the sp-carbon chains, while in $\alpha$-GDY/Pt 
the extra-bright contribution of the C atoms
at
the corners of the GDY porous structure emerges prominently.
Notably, the brightest atoms 
sit 
on top
of the substrate FCC sites and are less protruding toward vacuum.

The STM contrast is indeed mainly due to an electronic effect, being these atoms characterized by an excess of electrons.
The STM simulation of $\beta$-GDY on Au(111), 
obtained with identical setting as
adopted for $\alpha$-GDY, is overall brighter, due to the larger charge transfer toward the 
organic layer
(see below), and displays bright features in correspondence to the
carbon atoms at the corners of the starry structure, and a relatively intense contrast also along the sp-carbon chains.
Also in this case the dominant effect is 
electronic:
the height modulation is indeed 
relatively
small,
while the electronic charge accumulated on CC single bonds in the chain and at the corners of the structure is three times larger than the excess charge on triple C$\equiv$C bonds.
The STM image for $\beta$-GDY on Pt(111) is characterized by blurred spots in correspondence of the C atoms that interact 
most strongly with the underlying substrate, leading to bright features arising from a combination of structural deformation (atoms displaced toward vacuum) 
and electronic effect (charge transfer).
The more intense STM signal of GDY on Pt than on Au, reaching saturation on certain atoms in the adopted scale, reflects a larger local transfer of electrons on platinum (see details in section \ref{elec-prop}), giving a further evidence of the larger strength of the interaction with this substrate.

The total-energy
analysis confirms that $\alpha$-GDY is more strongly bound to Pt(111) than to Au(111).
Indeed, the adsorption energy
per carbon atom is  $-0.38$~eV on Au and
$-0.48$~eV on Pt (see Table~S1 in the Supporting Information for more details and definition of the relevant quantities).
It is worth noting that the energy required to strain and distort the GDY on Au ($0.078$~eV/atom) is about half that on
Pt ($0.14$~eV/atom), due to the larger distortion on Pt, and despite the smaller strain required to match the GDY to the Pt substrate spacing.
On the other hand, the pure-adhesion
energy contribution
is $-0.46$~eV/atom on Au and $-0.62$~eV on Pt.

The adsorption energy of $\beta$-GDY on both
substrates follows
similar
trends as for $\alpha$-GDY.
The energy cost of strain+distortion is slightly smaller ($0.02$~eV/atom on Au and $0.12$~eV/atom on Pt).
The total adsorption energies are the same as for $\alpha$-GDY
($-0.38$~eV on Au and $-0.48$~eV on Pt),
with pure-adhesion contributions
($-0.40$~eV/atom on Au and $-0.60$~eV/atom on Pt) that are also slightly smaller.

\subsubsection{Electronic properties}\label{elec-prop}

\begin{figure}
  \centering
  \includegraphics[width=0.8\linewidth]{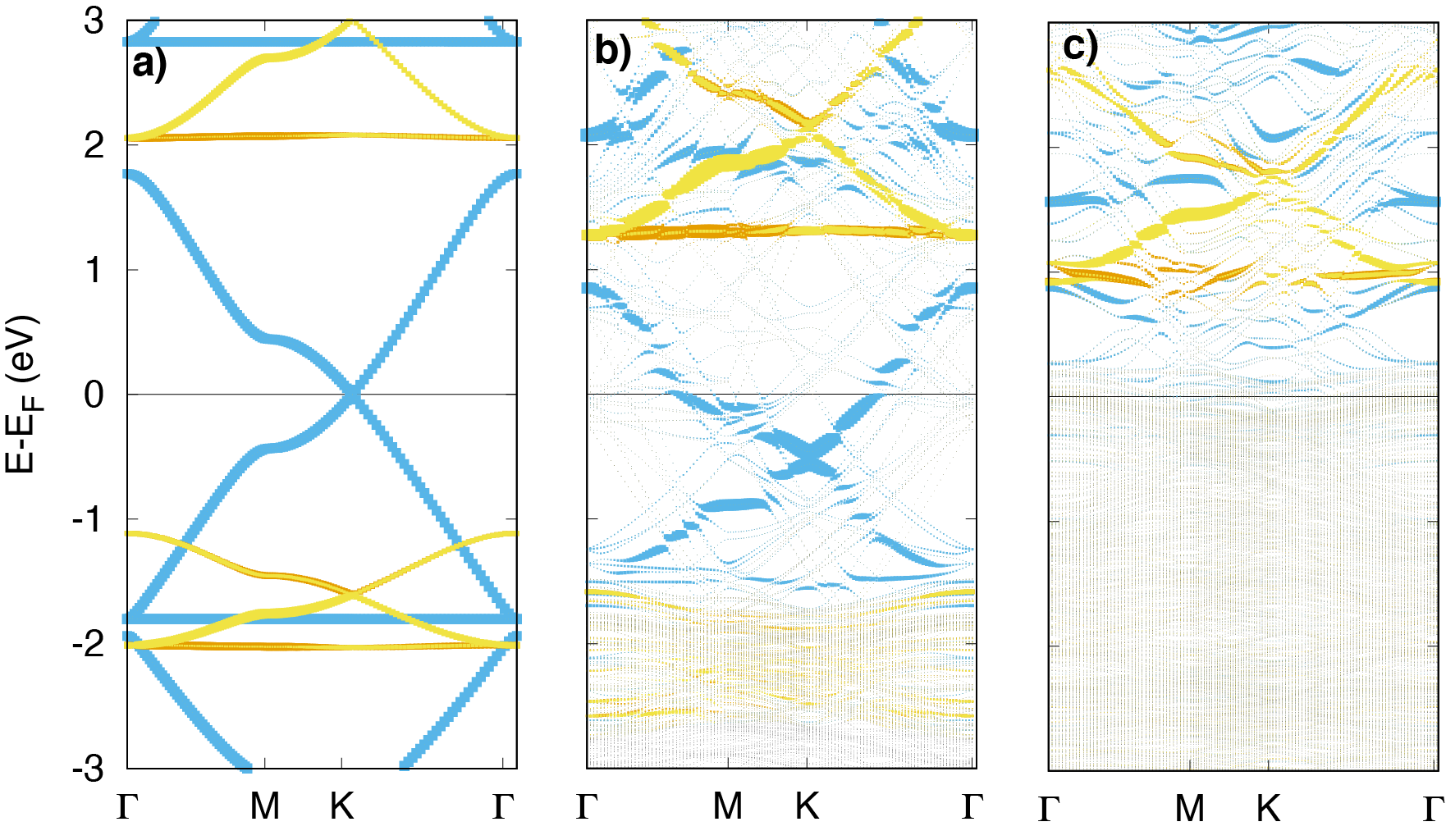}
  \caption{Band structure of $\alpha$-GDY in the following configurations: (a) freestanding; (b) adsorbed on Au(111); (c) adsorbed on Pt(111).
  The p$_z$ 
  weights
  are highlighted in blue and the p$_x$/p$_y$ ones in yellow.}
  \label{fig:bands-a}
\end{figure}

The stronger interaction of GDYs with Pt 
compared
to Au manifests itself also in the changes induced on the electronic properties in the adsorbed system.
The band structure of the $\alpha$-GDY on Au(111), reported in Figure~\ref{fig:bands-a}b, shows a shift of the Dirac cone
relative
to the freestanding case (Fig.\ref{fig:bands-a}a).
This $-0.5$~eV energy shift 
is accompanied by the opening of a small 
$0.1$~eV gap of (see also Figure~S2 in the SI)
In contrast,
the interaction with Pt(111) 
washes the Dirac-cone feature away: 
it mostly
disappears in the dense spectrum of the filled states of the underlying Pt surface.
In fact, by zooming around the Fermi level,
one can still
detect
a reminiscence of the Dirac cone just above the Fermi level, with upward and downward dispersing bands with relevant p$_z$ character (see Figure~S2 in SI). 

The dispersion of the p$_x$, p$_y$ bands is 
moderately
modified by the interaction with Au(111):
the main visible effects is indeed a downward energy shift mostly of the empty bands.
Differently, on Pt(111) the distorted GDY structure causes a substantial mixing with p$_z$ states, leading for example to a more dispersive behavior of the band near $+1$~eV,
which is otherwise flat in the freestanding GDY.

The opposite energy shift of the
filled bands
of $\alpha$-GDY on the two substrates is an effect of the opposite charge transfers occurring in the adsorbed GDYs.
The analysis of the Mulliken charges evidences a transfer of 0.018~electrons/C atom from Au(111) to $\alpha$-GDY and, conversely, of 0.01~electrons/C atom from the $\alpha$-GDY to the Pt substrate.

Notably, despite such an average electron transfer 
from $\alpha$-GDY to Pt(111),
certain C atoms in the structure 
receive
contributions with opposite sign.
In particular the C atoms
sitting 
in hollow positions are characterized by an excess of electron charge ($\sim 0.08$ electrons)
compared 
to the freestanding system, resulting in bright spots in STM images.
Differently, the
sp
C atoms 
forming the chains
are characterized by the largest reduction of charge ($\sim -0.07$ electrons).
On both substrates,
due to the interaction with the metal
and the ensuing charge transfer,
$\alpha$-GDY 
acquires
a 
frank
metallic character, as evidenced also by the density of states (DOS) reported in SI, Figure~S3.

\begin{figure}
  \centering
  \includegraphics[width=0.8\linewidth]{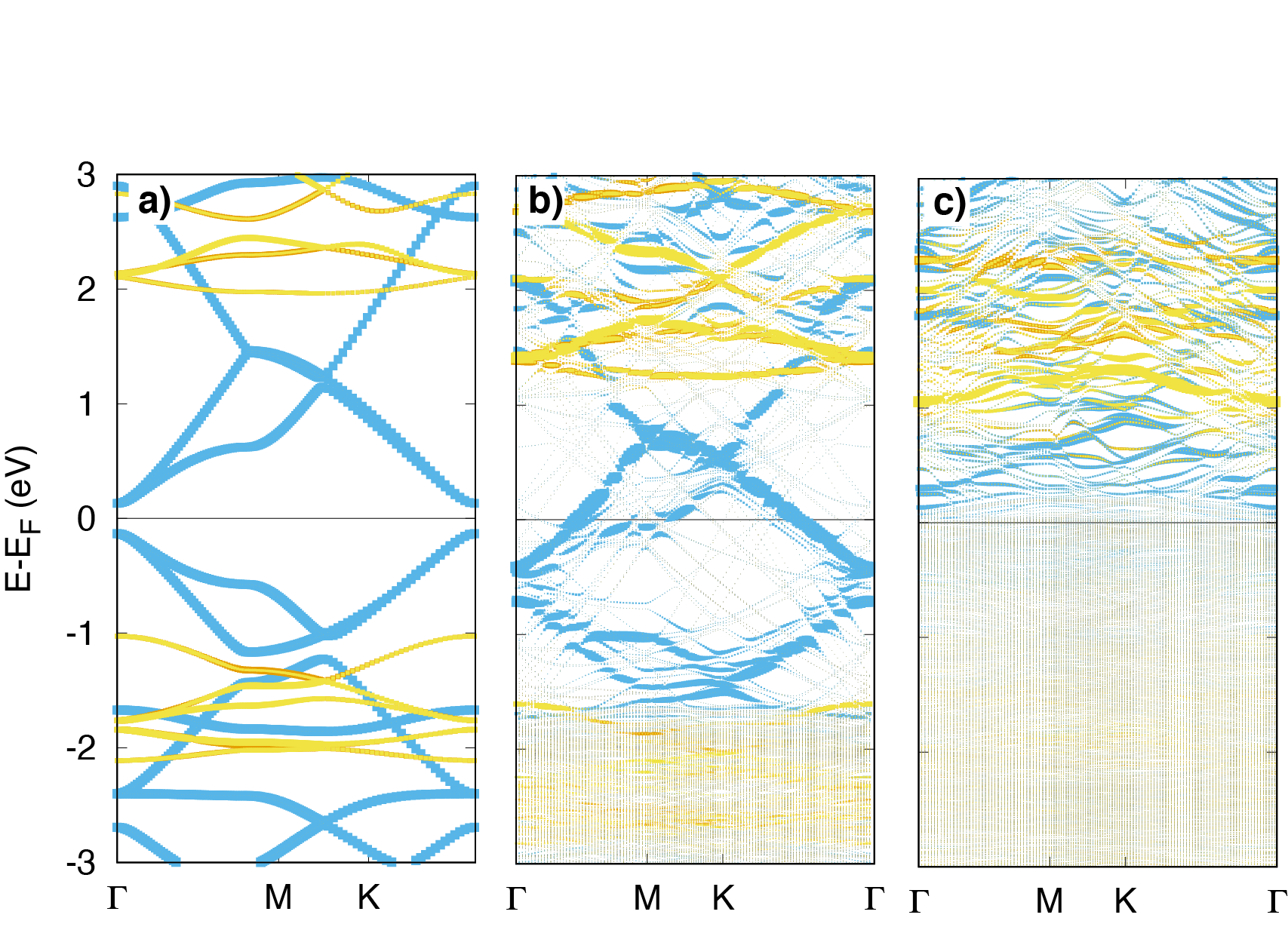}
  \caption{
  Same as Fig.~\ref{fig:bands-a}, but for 
$\beta$-GDY.
}
  \label{fig:bands-b}
\end{figure}

The freestanding $\beta$-GDY is a semiconductor with a gap of 0.26~eV, 
see Fig.~\ref{fig:bands-b}a.
When adsorbed on Au(111) $\beta$-GDY maintains the main features of 
its band structure, with the bands downshifted in energy by $\sim 0.5$~eV, as reported in Fig.~\ref{fig:bands-b}b.
Due to this shift, the $\beta$-GDY p$_z$ states cross the Fermi level
leading to a metallic character in the overlayer, as confirmed also by the DOS (see Figure~S3 in SI).

The effect of such charge transfer includes a displacement of the Dirac-cone-like feature that can be seen at $+1.3$~eV in the freestanding GDY, moving it closer to the Fermi level.

If a similar charge transfer was observed on a semiconducting/insulating substrate, it could be controlled via an electric field and exploited for switched electronic transport in a gated device setup.
Like for $\alpha$-GDY, the p$_x$ and p$_y$ bands are relatively weakly
modified upon adsorption on Au, apart for the downward energy shift  discussed above. On Pt(111), instead, the strong interaction and hybridization with the substrate states completely modifies the bands of the $\beta$-GDY overlayer (see also the broadening of the DOS in the SI, Figure~S3).

The effect of the structural deformation on the $\beta$-GDY bands is less
important than that of the hybridization with the metallic bands.
This is confirmed also by the bandstructure 
calculation of a freestanding $\beta$-GDY 
frozen in the distorted configuration induced by the 
Pt(111) substrate, which shows bands rather similar to those of the fully optimized (flat) freestanding structure (see SI, Figure~S1). 
Also in this case the changes of the states at the Fermi level due to the interaction with the substrate confer a metallic character to the carbon overlayer. 
The downshift of the conduction band of both GDYs below the Fermi level on the gold substrate and the strong hybridization with the platinum substrate states suggest that there is no vertical barrier for the flow of electrons at the interface, in analogy with what is found for $\gamma$-GDY on different metal surfaces \cite{Pan}.

The Mulliken charge analysis evidences a charge transfer of 0.013~electrons/C atom from Au surface
to the $\beta$-GDY, in agreement with the observed shift of the bands, while on Pt(111) a negligibly small average charge 
(0.0006 electrons/C atom) 
is transferred from the overlayer to the surface.
Like in the $\alpha$-GDY case,
the 
individual C atoms in the structure give contributions with opposite sign.
The maximum charge transfer on a single atom amounts to 0.06 electrons/C atom.

\subsubsection{Vibrational properties and simulated Raman spectra.}

Based on the validation obtained for h-GDY, 
we have applied the same computational approach to
$\alpha$- and $\beta$-GDY polymorphs and their interaction with both Au and Pt metal surfaces. 
The DFT-computed Raman spectra for these systems are reported in Figure~\ref{fig:struct}, including a comparison with the freestanding spectra.


Considering first the freestanding $\alpha$-GDY, three main bands are predicted by the simulations. The most intense one, located at 1950 cm$^{-1}$, is indeed the convolution of two intense bands, both of which are related to normal modes consisting in different Raman-active combinations of the ECC modes on the single diacetylenic $sp$-carbon units.
Another intense band is found at about 1420~cm$^{-1}$ and it is the convolution of two different bands (at 1420 and 1423~cm$^{-1}$ respectively), related to stretching of the C-C bonds between the sp$^2$ carbon atoms and the first sp carbon atoms of the diacetylenic units.
A weaker band is predicted as a convolution of the two bands at 784 and 785~cm$^{-1}$, and it is due mainly to CCC bending modes involving sp$^2$ carbon atoms. A further, and even weaker band is found at about 550 cm$^{-1}$, associated to linear bending modes of the sp-carbon domains.

\begin{figure}
  \centering
  \includegraphics[width=0.8\linewidth]{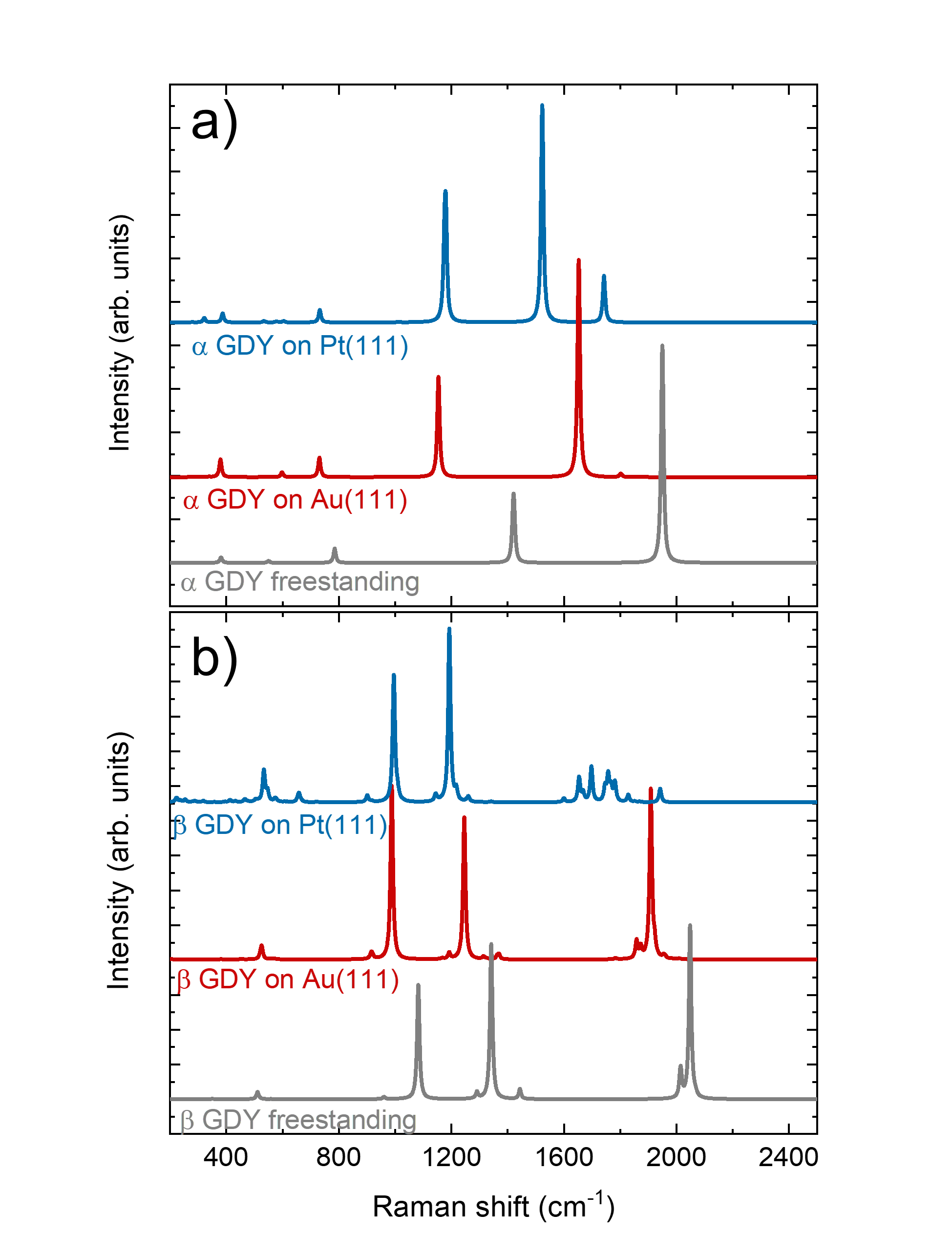}
  \caption{(a) DFT-simulated Raman spectra of $\alpha$-GDY freestanding and supported on Au(111) and on Pt(111).
  (b) DFT simulated Raman spectra of $\beta$-GDY freestanding and supported on Au(111) and on Pt(111). 
  }
  \label{fig:Raman-alfa-beta} 
\end{figure}

Our calculations predict
substantial modifications when $\alpha$-GDY interacts with Au and Pt surfaces.
The two strongest bands undergo
a large frequency shift.
The ECC bands at 1949~cm$^{-1}$ downshifts to 1652~cm$^{-1}$ upon interaction with Au and to 1523~cm$^{-1}$ upon interaction with Pt. 
Similarly, the 1421~cm$^{-1}$ band (i.e., CC stretching between $sp^2$ and $sp$ C atoms) moves down to 1154~cm$^{-1}$ on Au and to 1178~cm$^{-1}$ on Pt.
Furthermore, the bands associated to CCC bending modes shift from 784~cm$^{-1}$ to 730~cm$^{-1}$ on Au and 731~cm$^{-1}$ on Pt.
The interaction with the surface seems to affect more strongly the sp-carbon domains in the case of Pt, inducing a larger downshift of the ECC bands with respect to Au.
On the other hand, the geometry of GDY around the sp$^2$ carbon atoms is less affected on Pt, as demonstrated by the smaller downshift found for the 1421~cm$^{-1}$ band
compared to the Au case.
In addition to the bands already present in the
Raman spectrum of the
free-standing $\alpha$-GDY, when 
supported 
a further peak arises
at 1802~cm$^{-1}$ (weak) 
on Au(111)
and at 1742~cm$^{-1}$ on Pt(111), with an appreciable intensity.
These bands are associated to CC stretching vibrations involving the CC triple-bonds: these
lower-frequency
normal modes are the counterpart of the so-called “$\beta$” mode, observed and predicted for finite-length polyynes \cite{Casari_Nanoscale_2016, Milani_BeilsteinJ_2015}.
The larger intensities found on Pt indicate again that in this case the interaction with the metal more strongly affect the $sp$-carbon domains. 
In order to rationalize this spectral modulation, bond-lengths and BLA values have been analyzed.
BLA values on the $sp$-carbon domains change from 0.095~\AA\ in the free-standing, to 0.11~\AA\ on Au and to 0.078~\AA\ on Pt.
The lack of a clear trend in these values reveals that the downshifts predicted on metal surfaces cannot be simply related to the degree of conjugation usually leading to  
more equalized diacetylenic units \cite{milani2019structure}.
On the other hand,
the CC
bond-lengths
on the diacetylenic units become significantly longer upon interaction with the metal surfaces, with
elongations by
up to $0.066$~\AA.
This allows
us
to conclude that, also in this case, the interaction with Au and Pt surfaces gives more antibonding character to the CC bonds, weakening the bond strength, and thus resulting in lower C-C stretching wavenumbers.

Considering now the $\beta$-GDY polymorph, the freestanding
configuration exhibits four main bands
plus a few weaker 
ones.
The most intense band at 2048~cm$^{-1}$ is associated to ECC vibrations on the diacetylenic units.
Two other weaker contributions
forming
the band at 2015~cm$^{-1}$
are
also
associated to Raman active combinations of ECC vibrations on 
different diacetylenic units.
The other intense band at 1341 cm$^{-1}$ is mainly related to
the
stretching mode of the CC bond between 
adjacent
sp$^2$ carbon atoms. 
A fourth intense band is computed at 1082 cm$^{-1}$ due to CCC bending modes involving sp$^2$ carbon atoms.
As a consequence of the interaction with the Au surface, 
similar modifications
as for $\alpha$-GDY are observed.
All the intense bands display a consistent downshift in frequency, while keeping the same assignment in terms of normal modes of vibration. 
Such downshift is ascribed to a weakening of the CC bond of the network and a softening of the vibrational stretching force constants, as discussed for the $\alpha$-GDY 
case. In particular, at the corners of the structure bond lengths increase from $1.41$~\AA\ in the freestanding to $1.43$~\AA\ and $1.44$~\AA\ on Au and Pt, respectively. Triple and single bonds in the sp-carbon units elongate when moving from freestanding ($1.25$~\AA\ and $1.34$~\AA\ , respectively) to Au ($1.26$~\AA\ and $1.36$~\AA) and Pt ($1.30$~\AA\ and $1.35$~\AA) supported structure. However, the resulting BLA value increases when the overlayer is on Au (from $0.097$~\AA\ to $0.099$~\AA) while decreases on Pt (from $0.097$~\AA\ to $0.050$~\AA). A complex interplay between $\pi$-electron conjugation and structural parameters occurs also in this case, showing that changes in conjugation can have a non trivial effect on the BLA \cite{SERAFINI-CArbon2021}.

Conversely, when interacting with Pt, the Raman spectrum shows a more complicated pattern, in particular in the region between 1500 and 2000~cm$^{-1}$ where many bands of comparable intensities are predicted, spanning an extended range of wavenumbers. 
This is due to the fact that, as shown in Figure~\ref{fig:struct}, the structure of $\beta$-GDY is strongly deformed upon interaction with the Pt surface, and some of the diacetylenic units
deviate significantly
from linearity, showing bending angles of about 160$^{\circ}$. 
As a consequence of this structural deformation, the normal modes in this region are not anymore acting as a collective CC stretching mode delocalized
across
the whole sp-carbon units as
in the freestanding crystal, but are now
localized 
on different
CC stretching vibrations
along the distorted sp-carbon units, each mode usually involving mainly one or a few CC bonds.
Hence a multiplicity of bands arises due to the variety of different CC bond stretching,
resulting from
the large structural deformation induced by the Pt surface.

Based on these results, Raman spectroscopy
confirms its
usefulness for the characterization of GDYs and other hybrid sp-sp$^2$ carbon materials, providing important information on their structural properties and on the effects of the interaction with metal surfaces.
The strong sensitivity of the Raman features on small changes of the molecular structure allows one to use this technique to investigate subtle effects, related to both the inherent structure of the carbon 2D network and to its interactions with the substrate.

\section{Conclusions}

Building on recent experimental and theoretical results on h-GDY
\cite{rabia2020ANM},
we 
unveil the mechanisms underlying the changes of electronic and vibrational properties
of different GDYs 
induced by
the presence of a metal substrate. 
The present work aims at predicting properties that 
will be easily measured
as soon as
the so-far-scarcely-characterized
$\alpha$-GDY and $\beta$-GDY on Au and Pt are synthesized and studied experimentally.
In particular, we show that the Dirac cones
that characterize
the semimetallic free-standing $\alpha$-GDY undergo an energy shift in the 
interaction with
an Au substrate, while they completely disappear in the case of Pt.
The $\beta$-GDY polymorph, instead, which is semiconducting in its free-standing form, becomes metallic when deposited on both Au and Pt surfaces.
Overall, this analysis confirms the possibility to tune the electronic properties of the GDYs by appropriately choosing the substrate.
In particular, the choice of Au(111) allows one to preserve the main features of the GDY band structure while substrates with strongly-interacting d-states at the Fermi level, such as Pt(111), induce strong deformation of both the structural and electronic properties of the 2D carbon network.

Such important changes in the electronic structure are accompanied by 
dramatic
effects on the predicted Raman spectra. 
Since full Raman simulations for extended GDYs on metal substrates are known to suffer from severe limitations, 
in the present work we proposed a
simplified
approach: we keep the geometry of the adsorbed GDY, and remove the metal surface altogether:
this approach retains
those effects on the Raman spectra which can be ascribed to the {\em structural changes} that the 
GDYs undergo in the presence of the
metallic substrate, while leaving out charge-transfer and substrate-GDY
covalency effects.

We successfully validate 
this approach on the Au(111)-synthesized h-GDY case, whose
experimental
Raman spectra 
are available.
By applying this method to 
$\alpha$- and $\beta$-GDY, we observe Raman features which are consistent with the structural modifications induced by the presence of the metal substrate such as frequency shifts and splittings of the ECC normal modes, thus 
confirming
that Raman spectroscopy can provide a wealth of information for GDYs on metals.  

The importance of  sp-sp$^2$ bidimensional overlayers, 
such as GDYs,
has been recently
highlighted
in view of their potential  technological applications, which could exploit their peculiar functional properties \cite{Bryce-review2021}.
In this framework, predictions of the
substrate effect are crucial, since 
substrates themselves
provide additional degrees of freedom to the engineering of novel 2D materials.
A clear understanding and control of the GDY-substrate interactions, will
provide us with a suitable spectrum of choices of both the GDY polymorph 
and the underlying substrate, thus
opening the way toward a controlled tuning and engineering of the properties of the organic overlayer.

\section{Acknowledgments}
A.M., F.T., and C.S.C.\ acknowledge funding from the European Research Council (ERC) under the European Union’s Horizon 2020 research and innovation program ERC Consolidator Grant (ERC CoG2016 EspLORE grant agreement no.\ 724610, website: www.esplore.polimi.it). 
S.A., G.F., N.M., and G.O.\ acknowledge CINECA for the use supercomputing facilities under the agreement with UNITECH-INDACO. 
N.M.\ acknowledges support from the grant PRIN2017 UTFROM of the Italian Ministry of University and Research.

\section*{References}
\bibliographystyle{iopart-num}

\begin{thebibliography}{10}
\expandafter\ifx\csname url\endcsname\relax
  \def\url#1{{\tt #1}}\fi
\expandafter\ifx\csname urlprefix\endcsname\relax\def\urlprefix{URL }\fi
\providecommand{\eprint}[2][]{\url{#2}}

\bibitem{Hirsch_NatMat_2010}
Hirsch A {2010} {\em {Nat. Mater.}\/} {\bf {9}} {868--871}

\bibitem{Sakamoto-AdvMat2019-rev}
Sakamoto R, Fukui N, Maeda H, Matsuoka R, Toyoda R and Nishihara H 2019 {\em
  Adv. Mater.\/} {\bf 31} 1804211

\bibitem{Liu21}
Liu Y, Xue Y, Yu H, Hui L and Huang B~andLi Y 2021 {\em Adv. Funct. Mater.\/}
  {\bf 31} 2010112

\bibitem{Zou20}
Zou H, Rong W, Wei S, Ji Y and Duan L 2020 {\em Proc. Natl. Acad. Sci. USA\/}
  {\bf 117} 29462--29468

\bibitem{Yu21}
Yu H, Xue Y, Hui L, Zhang C, Fang Y, Liu Y, Chen X, Zhang D, Huang B and Li Y
  2021 {\em Natl. Sci. Rev.\/} {\bf 8} 8

\bibitem{baughman1987}
Baughman R~H, Eckhardt H and Kertesz M 1987 {\em J. Chem. Phys.\/} {\bf 87}
  6687--6699

\bibitem{Huang_2018}
Huang C, Li Y, Wang N, Xue Y, Zuo Z, Liu H and Li Y 2018 {\em Chem. Rev.\/}
  {\bf 118} 7744--7803

\bibitem{Casari_Nanoscale_2016}
Casari C~S, Tommasini M, Tykwinski R~R and Milani A {2016} {\em {Nanoscale}\/}
  {\bf {8}} {4414--4435}

\bibitem{Casari_MRSComm_2018}
Casari C~S and Milani A {2018} {\em {MRS Commun.}\/} {\bf {8}} {207--219}

\bibitem{Serafini-Proserpio-JPCC2021}
Serafini P, Milani A, Proserpio D~M and Casari C~S 2021 {\em J. Phys. Chem.
  C\/} {\bf 125}(33) 18456--18466

\bibitem{Gao_2019}
Gao X, Liu H, Wang D and Zhang J 2019 {\em Chem. Soc. Rev.\/} {\bf 48}(3)
  908--936

\bibitem{ChenZhi-AnnDerPhys2017}
Chen Z, Molina-Jirón C, Klyatskaya S, Klappenberger F and Ruben M 2017 {\em
  Ann. Phys.\/} {\bf 529} 1700056

\bibitem{Jia_2017}
Jia Z, Li Y, Zuo Z, Liu H, Huang C and Li Y 2017 {\em Accounts of Chemical
  Research\/} {\bf 50} 2470--2478

\bibitem{Li_NanoEn_2018}
Li M, Wang Z~K, Kang T, Yang Y, Gao X, Hsu C~S, Li Y and Liao L~S {2018} {\em
  {Nano Energy}\/} {\bf {43}} {47--54}

\bibitem{Zuo_2018}
Zuo Z, Wang D, Zhang J, Lu F and Li Y 2019 {\em Adv. Mater.\/} {\bf 31} 1803762

\bibitem{Li_2010_GDY}
Li G, Li Y, Liu H, Guo Y, Li Y and Zhu D 2010 {\em Chem. Commun.\/} {\bf
  46}(19) 3256--3258

\bibitem{Li_ChemSocRev_2014}
Li Y, Xu L, Liu H and Li Y {2014} {\em {Chem. Soc. Rev.}\/} {\bf {43}}
  {2572--2586}

\bibitem{Ivanovskii_ProgSolidStChem_2013}
Ivanovskii A~L {2013} {\em {Progr. Solid State Ch.}\/} {\bf {41}} {1--19}

\bibitem{serafini2020raman}
Serafini P, Milani A, Tommasini M, Castiglioni C and Casari C~S 2020 {\em Phys.
  Rev. Mater.\/} {\bf 4} 014001

\bibitem{Zhang_JPCC2016-graphyne}
Zhang S, Wang J, Li Z, Zhao R, Tong L, Liu Z, Zhang J and Liu Z {2016} {\em {J.
  Phys. Chem. C}\/} {\bf {120}} {10111--10720}

\bibitem{Milani_BeilsteinJ_2015}
Milani A, Tommasini M, Russo V, Bassi A~L, Lucotti A, Cataldo F and Casari C~S
  {2015} {\em {Beilstein J. Nanotechnol.}\/} {\bf {6}} {480--491}

\bibitem{zhang2012homo}
Zhang Y~Q, Kep{\v{c}}ija N, Kleinschrodt M, Diller K, Fischer S, Papageorgiou
  A~C, Allegretti F, Bj{\"o}rk J, Klyatskaya S, Klappenberger F, Ruben M and
  Barth J~V 2012 {\em Nat. Commun.\/} {\bf 3} 1286

\bibitem{Sun_JACS_2016}
Sun Q, Cai L, Wang S, Widmer R, Ju H, Zhu J, Li L, He Y, Ruffieux P, Fasel R
  and Xu W 2016 {\em J. Am. Chem. Soc.\/} {\bf {138}} {1106--1109}

\bibitem{Sun_AngewChem_2017}
Sun Q, Tran B~V, Cai L, Ma H, Yu X, Yuan C, Stohr M and Xu W {2017} {\em
  {Angew. Chem. Int. Ed.}\/} {\bf {56}} {12165--12169}

\bibitem{sun2016}
Sun Q, Cai L, Ma H, Yuan C and Xu W 2016 {\em ACS Nano\/} {\bf 10} 7023--7030

\bibitem{Fan20152484}
Fan Q, Gottfried J and Zhu J 2015 {\em Acc. Chem. Res.\/} {\bf 48} 2484--2494

\bibitem{klappenberger2015surface}
Klappenberger F, Zhang Y~Q, Bj{\"o}rk J, Klyatskaya S, Ruben M and Barth J~V
  2015 {\em Acc. Chem. Res.\/} {\bf 48} 2140--2150

\bibitem{Shu_NatComm_2018}
Shu C~H, Liu M~X, Zha Z~Q, Pan J~L, Zhang S~Z, Xie Y~L, Chen J~L, Yuan D~W, Qiu
  X~H and Liu P~N {2018} {\em {Nat. Commun.}\/} {\bf {9}} 2322

\bibitem{Pan}
Pan Y, Wang Y, Wang L, Zhong H, Quhe R, Ni Z, Ye M, Mei W~N, Shi J, Guo W, Yang
  J and Lu J {2015} {\em {Nanoscale}\/} {\bf {7}}

\bibitem{rabia2020ANM}
Rabia A, Tumino F, Milani A, Russo V, Li~Bassi A, Bassi N, Lucotti A, Achilli
  S, Fratesi G, Onida G, Manini N, Sun Q, Xu W and Casari C~S 2020 {\em ACS
  Appl. Nano Mater.\/} {\bf 3} 12178--12187

\bibitem{Yan}
Yan H, Yu P, Han G, Zhang Q, Gu L, Yi Y, Liu H, Li Y and Mao L 2019 {\em Angew.
  Chem. Int. Ed.\/} {\bf 58} 746--750

\bibitem{rabia2019scanning}
Rabia A, Tumino F, Milani A, Russo V, Li~Bassi A, Achilli S, Fratesi G, Onida
  G, Manini N, Sun Q, Xu W and Casari C~S 2019 {\em Nanoscale\/} {\bf 11}(39)
  18191--18200

\bibitem{Borrelli-Angwchemie2021}
Borrelli M, Querebillo C, Dominik~L P, Tao W, Alberto M, Carlo~S C, Ly H~K, He
  F, Hou Y, Christof N, Andrey T, Sun H, Inez~M W and Xinliang F 2021 {\em
  Angew. Chem. Int. Ed.\/} {\bf 60} 2--8

\bibitem{Galeotti-NatMat2020}
Galeotti G, Marchi F~D, Hamzehpoor E, MacLean O, Rao M~R, Chen Y, Besteiro L,
  Dettmann D, Ferrari L, Frezza F, Sheverdyaeva P, Liu R, Kundu A, Moras P,
  Ebrahimi M, Gallagher M, Rosei F, Perepichka D and Contini G 2020 {\em Nat.
  Mater.\/} {\bf 19} 874–880

\bibitem{PBE}
Perdew J~P, Burke K and Ernzerhof M 1996 {\em Phys. Rev. Lett.\/} {\bf 77}(18)
  3865--3868

\bibitem{Grimme}
Grimme S 2006 {\em J. Comput. Chem.\/} {\bf 27}(15) 1787--1799

\bibitem{Sole02}
Soler J~M, Artacho E, Gale J~D, Garc{\'\i}a A, Junquera J, Ordej\'on P and
  S\'anchez-Portal D 2002 {\em J. Phys.: Condens. Matter\/} {\bf 14} 2745

\bibitem{Tersoff}
Tersoff J and Hamman D~R 1985 {\em Phys. Rev. B\/} {\bf 31} 805

\bibitem{Adamo-BaroneJPC1999-PBE0}
Adamo C and Barone V 1999 {\em J. Chem. Phys.\/} {\bf 110} 6158--6170

\bibitem{TommasiniJPCA2007}
Tommasini M, Fazzi D, Milani A, Del~Zoppo M, Castiglioni C and Zerbi G 2007
  {\em J. Phys. Chem. A\/} {\bf 111} 11645--11651

\bibitem{Castelli2012}
Castelli I, Ferri N, Onida G and Manini N {2012} {\em { J. Phys.: Condens.
  Matter }\/} {\bf {24}} {104019}

\bibitem{milani2019structure}
Milani A, Barbieri V, Facibeni A, Russo V, Bassi A~L, Lucotti A, Tommasini M,
  Tzirakis M~D, Diederich F and Casari C~S 2019 {\em Scientific reports\/} {\bf
  9} 1--10

\bibitem{SERAFINI-CArbon2021}
Serafini P, Milani A, Tommasini M, Bottani C~E and Casari C~S 2021 {\em
  Carbon\/} {\bf 180} 265--273

\bibitem{Bryce-review2021}
Bryce M~R 2021 {\em J. Mater. Chem. C\/} {\bf Advance Article}({})

\end{thebibliography}

\providecommand{\newblock}{}

\end{document}